# A Review of Stable, Traversable Wormholes in $f(R)$ Gravity Theories

Ramesh Radhakrishnan [1,2], Patrick Brown[1,2], Jacob Matulevich[1,2], Eric Davis[1,2], Delaram Mirfendereski[1,3], and Gerald Cleaver[1,2]

[1]*Early Universe, Cosmology and Strings (EUCOS) Group, Center for Astrophysics, Space Physics and Engineering Research (CASPER)*
[2]*Department of Physics, Baylor University, Waco, TX 76798, USA*
[3]*Department of Physics and Astronomy, The University of Texas Rio Grande Valley (UTRGV), USA*

September 3, 2024

**Abstract**

It has been proven that in standard Einstein gravity, exotic matter (i.e., matter violating the pointwise and averaged Weak and Null Energy Conditions) is required to stabilize traversable wormholes. Quantum field theory permits these violations due to the quantum coherent effects found in any quantum field . Even reasonable classical scalar fields violate the energy conditions. In the case of the Casimir effect and squeezed vacuum states, these violations have been experimentally proven. It is advantageous to investigate methods to minimize the use of exotic matter. One such area of interest is extended theories of Einstein gravity. It has been claimed that in some extended theories, stable traversable wormholes solutions can be found without the use of exotic matter. There are many extended theories of gravity, and in this review paper, we first explore $f(R)$ theories and then explore some wormhole solutions in $f(R)$ theories, including Lovelock gravity and Einstein Dilaton Gauss–Bonnet (EdGB) gravity. For completeness, we have also reviewed 'Other wormholes' such as Casimir wormholes, dark matter halo wormholes, thin-shell wormholes, and Nonlocal Gravity (NLG) wormholes, where alternative techniques are used to either avoid or reduce the amount of exotic matter that is required.

# 1   Introduction

The first study of wormhole physics was done by Ludwig Flamm in 1916 [1] during his research into the Schwarzschild solution to the Einstein field equations. The next solution resembling a wormhole, called the "Einstein–Rosen bridge" [2], was an idea that arose during the investigation of blackhole spacetimes by Einstein and Rosen in 1935. They discovered that, at least theoretically, it was possible for a blackhole surface to act as a bridge that connected to a remote patch of spacetime. The putative wormhole in the Einstein–Rosen bridge is colocated with the blackhole's singularity, so it is not traversable. The surface of the blackhole is the event horizon, which cannot be a wormhole.

A good review of the historic development of research into traversable wormholes can be found in [3]. A more recent concise summary of wormholes in extended theories of gravity can be found in [4]. In 1957, Misner and Wheeler [5] first introduced the term "wormhole" during their analysis of topological issues in General Relativity (GR). They extensively analyzed the Riemannian geometry of manifolds of nontrivial topology. This is where the phrase "physics is geometry" arose. Wheeler also suggested that the geometry of spacetime might be constantly fluctuating, and it may induce a change in topology to form microscopic wormholes. The first traversable wormhole called the "Ellis drainhole" was proposed by both Ellis and independently by Bronnikov in 1973 [6–8]. The "Ellis drainhole" spacetime is a static, spherically symmetric solution of the Einstein field equations in a vacuum, and it includes a scalar field $\phi$ minimally coupled to the spacetime geometry.

The concept of traversable wormholes, which allow inter- and intrauniverse travel by humans, was first introduced by Morris and Thorne in their classic 1988 paper [9]. It is well known that the Weak and Null Energy Conditions (WECs and NECs) [10] must be violated by the stress–energy tensor as a minimum for a stable human traversable wormhole. The other two energy conditions are the Strong Energy Condition (SEC) and the Dominant Energy Condition (DEC) [11]. The violation of the NEC is needed due to the flare-out condition requirement, i.e., the throat should open up outward as a human travels through the wormhole [9, 10]. Quantum field theory permits these violations due to the quantum coherent effects found in any quantum field [12]. The traversable wormhole solutions have geometries that allow for closed timelike curves and "effective" superluminal travel without surpassing the speed of light locally [3]. There is a claim [13] in the published literature about stable traversable wormholes that can be constructed within the framework of Einstein's General Relativity (henceforth called 'Einstein Gravity' in this paper) but without the need for exotic matter. This claim has been refuted by other authors [14]. Some of the suggested reasons in [14] are the following: (i) possible error in calculations; (ii) failing to check for violations at the throat, such as divergence of the inverse of the metric; and (iii) failing to check for discontinuities in the exterior curvature in the vicinity of the throat based on the thin-shell formalism. Some simple examples of traversable wormholes such as the polyhedral wormholes that do not require spherical symmetry were given by Matt Visser in 1989 [15, 16]. These wormholes also require the presence of exotic matter.

## 1.1 Motivation and Equation of State

There are many attempts to use various modified gravity theories to check the existence of stable traversable wormholes in these theories. In many of these studies [17–19], the Morris–Thorne (MT) traversable wormhole has been used to see the effects of a modified gravity background on the stability of the wormhole. One of the main motivations for studying wormholes in modified gravity theories is to resolve the problem of the need for exotic matter to stabilize wormholes in GR. Modified gravity theories are also used to construct viable cosmological models of the Universe and to explain singularities encountered in cosmology. In modified gravity theories, the stress energy tensor is replaced by an effective stress energy tensor that contains curvature terms of higher order introduced due to modifications to GR. $f(R)$ gravity [20, 21] is one such leading theory, and the review of wormholes in $f(R)$ gravity is one of the main goals of this paper.

While evaluating wormholes in GR and various modified theories, the choice of an equation of state (EOS) becomes critical. The EOS is an equation that relates the radial pressure $p_r(r)$ and tangential pressure $p_t(r)$ to the energy density $\rho(r)$. One of the most common EOSs is $p(r) = \omega\rho(r)$. The motivation for using such an equation comes from cosmology. Observation of the accelerated expansion of the Universe can be explained by an EOS with $\omega < -\frac{1}{3}$ [22]. For the case where $\omega < -1$, the EOS is known as the phantom energy EOS. Phantom fluid violates the NEC and is a good candidate to be used in the study of wormholes [23] in various modified gravity theories. According to a study [24], if a source with such phantom fluid dominates the cosmic expansion, the Universe may end up in a Big Rip singularity in which the phantom energy rips apart the galaxies, solar systems, planets, and, eventually, the molecules, atoms, nuclei, and nucleons that we are made of, leading to the death of the Universe.

## 1.2 Astronomical Observational Signatures of Traversable Wormholes

The presence of naturally occurring negative energy regions in space is predicted to produce a unique signature corresponding to lensing, chromaticity, and intensity effects in micro- and macrolensing events on galactic and extragalactic/cosmological scales [25–30]. It has been shown that these effects provide a specific signature that allows for discrimination between ordinary (positive mass–energy) and negative energy lenses via the spectral analysis of astronomical lensing events. The theoretical modeling of negative energy lensing effects has led to intense astronomical searches for naturally occurring traversable wormholes in the universe. Computer model simulations and comparison of their results with recent satellite observations of gamma ray bursts (GRBs) have shown that putative negative energy (i.e., traversable wormhole) lensing events very closely resemble the main features of some GRBs.

When background light rays strike a negative energy lensing region, they are swept out of the central region, thus creating an umbra region of zero intensity [25]. At the edges of the umbra, the rays accumulate and create a rainbow-like caustic with enhanced light intensity. The lensing of a negative energy region is not analogous to a diverging lens because, in certain circumstances, it can produce more light enhancement than does the lensing of an equivalent positive mass–energy region [25]. Real background sources in lensing events can have nonuniform brightness distributions on their surfaces and a dependency of their emission on the observed frequency. These complications can result in chromaticity effects, i.e., in spectral changes induced by differential lensing during the event. The quantification of such effects is quite lengthy, somewhat model dependent, and with recent application only to astronomical lensing events.

The rest of this paper is organized as follows: In Section 2, we provide an overview of various modified gravity theories with just enough background necessary to analyze traversable wormholes in such theories. In Section 3, we review the basics of Morris–Thorne (MT) wormhole stabilization in GR. We end Section 3 with a summary of the steps used to analyze MT wormholes in modified gravity theories. In Section 4, which forms the core of this paper, we review the analysis of wormhole stabilization in various f(R) theories. In Section 5, we review some closely related topics such as Casimir wormholes, thin-shell wormholes, natural dark matter halo wormholes, and wormholes in nonlocal theories of gravity (NLGs). We end this paper with a summary of key discussion points in this paper.

## 2 Modified Gravity Theories

There have been efforts to construct stable traversable wormholes in $f(R)$ theories, including Lovelock gravity. In this section, we give a brief overview of the basics of these theories to provide us with enough background to explore in the next section the properties of traversable wormholes.

### 2.1 An Overview of Modified Gravity Theories

The various modified gravity theories that we survey here are $f(R)$ theories, $f(T)$ theories, $f(R, \mathscr{T})$ theories, $f(G)$ theory, $f(R, L_m)$ theory, $f(Q)$ theory, $f(Q, T)$ theory, Lovelock gravity (a special case of $f(R)$ theory), Einstein–Gauss–Bonnet (EGB) gravity (another special case of $f(R)$ theory), Brans–Dicke theory, and Kaluza-Klein (KK) theory.

Lovelock gravity and EGB gravity are $f(R)$ theories that include higher-order curvature terms, and they are also applicable to higher dimensional spacetimes. For example, in EGB gravity [10], the authors replace the 2-sphere in the MT wormhole with an (n-2)-sphere, and thus the MT wormhole line element is modified as follows:

$$ds^2 = -e^{2\phi(r)}dt^2 + \frac{dr^2}{1 - b(r)/r} + r^2 d\Omega_{n-2}^2. \tag{1}$$

The standard Einstein–Hilbert action of GR is

$$S = \int R\sqrt{-g}\, d^4x. \tag{2}$$

In $f(R)$ theories [31], the Ricci scalar $R$ is replaced by a function of the Ricci Scalar $f(R)$ as follows:

$$S = \int f(R)\sqrt{-g}\, d^4x, \tag{3}$$

where $f(R)$ is a function of the Ricci scalar, and g is the determinant of the metric. Examples of commonly used functions $f(R)$ are $f(R) = R$, $f(R) = R + \alpha R^n$, and $f(R) = R + \alpha e^{(\beta R)}$.

In teleparallel gravity [32, 33], the Ricci scalar $R$ is replaced by the torsion scalar $T$ as follows:

$$S = \int T\sqrt{-g}\, d^4x. \qquad (4)$$

In GR, we assume spinless particles to follow the geodesic of the underlying spacetime, and hence, we have only $R$ in the action and no $T$. In teleparallel gravity, $T$ replaces $R$. This interpretation holds true in the case where we see teleparallel gravity as a gauge theory [34] for the translation group.

The $f(T)$ gravity [35] is an additional modification to teleparallel gravity, in which $T$ is replaced by the torsion function—which is a function of $T$, namely $f(T)$, as follows:

$$S = \int f(T)\sqrt{-g}\, d^4x, \qquad (5)$$

and we can obtain the modified field equations by varying the action $S$ with respect to the metric in the same way it is done in Section 2.2 of this paper for $f(R)$ theory.

$T$ is a fundamental geometric quantity in the context of theories of gravity that involve spacetime torsion. Unlike GR, where spacetime curvature plays a central role in describing gravity, theories that incorporate torsion consider the twisting or nonmetricity of spacetime, and the torsion scalar is a measure of this twisting. The torsion scalar is given by the equation

$$T = S_{\alpha\beta\mu}T^{\alpha\beta\mu}, \qquad (6)$$

where $S_{\alpha\beta\mu}$ is the contorsion tensor, which is the difference between the affine (Levi–Civita) connection's components and the Weitzenböck connection's components. The contorsion tensor quantifies nonmetricity, and it describes how spacetime is twisted. The indices $\alpha$, $\beta$, and $\mu$ refer to spacetime coordinates. In teleparallel gravity theory, the teleparallel connection is used instead of the metric affine connection that is used in GR.

In the $f(R, \mathscr{T})$ theory, $\mathscr{T}$ is the trace of the energy–momentum tensor $T_{\mu\nu}$. Similar to $f(R)$ theories [17], the presence of $f(R, \mathscr{T})$ in the action leads to changes in gravitational dynamics as compared to Einstein gravity. These modifications can have implications on the behavior of gravitational fields in various contexts. They also have consequences in cosmology and gravitational lensing. An example of an $f(R, \mathscr{T})$ function used to stabilize a wormhole is

$$f(R, \mathscr{T}) = R + \alpha R^2 + \lambda \mathscr{T}, \qquad (7)$$

where $\alpha$ and $\lambda$ are constants. In $f(R, \mathscr{T})$ theory, wormhole solutions with normal matter are feasible when appropriate shape functions are used. The coupling parameters $\alpha$ and $\lambda$ in the action of $f(R, \mathscr{T})$ theory play an important role in accommodating the composition of matter. According to [17], when $\alpha < 0$, wormholes exist in the presence of exotic matter, and when $\alpha > 0$, wormholes exist even in the absence of exotic matter.

In Brans–Dicke theory [36], a scalar field $\phi$ is introduced to modify the standard Einstein–Hilbert action as follows [37]:

$$S = \int [\phi R - \frac{\omega}{\phi}\phi_{;\mu}\phi^{;\mu} + \mathscr{L}_m]\sqrt{-g}\, d^4x, \qquad (8)$$

where $\phi$ is the Brans–Dicke scalar field, $\phi_{;\mu}$ is the covariant derivative of $\phi$, and $\omega$ is a coupling constant that couples the Brans–Dicke scalar field $\phi$ with the gravitational field. $\mathscr{L}_m$ is the Lagrangian density for the matter field(s).

The field equations in Brans–Dicke theory can be obtained as always by varying the action. These field equations are modified as compared to Einstein's field equations in GR, and they can be expressed in terms of the scalar field $\phi$ and its derivatives, as well as the metric tensor, the curvature tensors, and the scalars. The value of the coupling parameter $\omega$ affects the behavior of the theory, and it can influence the stability of wormholes.

Kaluza–Klein (KK) theory [38] extends the usual four-dimensional spacetime of GR to include one or more extra dimensions. These extra dimensions are compactified or 'curled up' so small that they are not perceptible on macroscopic scales. The total spacetime is a product of the usual four-dimensional spacetime and the compactified extra dimensions.

In KK theory, the metric tensor describes the geometry of the higher-dimensional spacetime. It has components corresponding to both the usual four dimensions denoted by $\mu$ and $\nu$ and the extra dimensions denoted by $a$ and $b$. The metric tensor can be decomposed into a four-dimensional part and an extra-dimensional part. The extra-dimensional part manifests as an additional vector or scalar fields, and these fields are typically associated with electromagnetic interactions. This process of decomposing the higher dimensional metric tensor field is called dimensional reduction.

The action of KK theory after dimensional reduction is

$$S = \int [R_{(4)} - F^{ab}F_{ab} - \mathscr{L}_m]\sqrt{-g}\, d^4x, \tag{9}$$

where $d^4x$ is the four-dimensional spacetime volume element, $R_{(4)}$ is the four-dimensional scalar curvature, $F^{ab}$ is the electromagnetic field strength tensor, and $\mathscr{L}_m$ is the usual Lagrangian density for the matter field(s).

There are currently three sets of geometric theories of gravity. The first one is the general theory of relativity based on curvature. The second is teleparallel gravity based on torsion (T), as we have seen in $f(T)$ theory. The third set of geometric theories is based on nonmetricity (Q), as described here. The origin of $f(Q)$ theory is from 'symmetric teleparallel gravity', which is based on the nonmetricity scalar Q. Nonmetricity is defined as the covariant derivative of the metric tensor $g_{\mu\nu}$, i.e., $Q_{\alpha\mu\nu} \equiv \nabla_\alpha g_{\mu\nu}$. It vanishes in the case of Riemannian geometry, and it can be used to study non-Riemannian spacetimes. $f(Q)$ gravity has inspired research in blackholes, wormholes, and cosmology. In cosmology, $f(Q)$ models can be used to explain phenomena related to both early and late time cosmology [39], without dark energy, dark matter, or the inflation field. The action of $f(Q)$ gravity is given by [40]

$$S = \int [-\frac{1}{16\pi G}f(Q) + \mathscr{L}_m]\sqrt{-g}d^4x. \tag{10}$$

Here, as usual, g is the determinant of the metric $g_{\mu\nu}$, and $\mathscr{L}_m$ is the matter Lagrangian density. $f(Q)$ is an arbitrary function of the nonmetricity scalar $Q$ [41] given by

$$Q = -\frac{1}{4}Q_{\alpha\beta\gamma}Q^{\alpha\beta\gamma} + \frac{1}{2}Q_{\alpha\beta\gamma}Q^{\gamma\beta\alpha} + \frac{1}{4}Q_\alpha Q^\alpha - \frac{1}{2}Q_\alpha \tilde{Q}^\alpha, \tag{11}$$

where $Q_\alpha \equiv Q^\mu_{\alpha\mu}$ and $\tilde{Q}_\alpha \equiv Q^{\mu\alpha}_\mu$ are two independent traces of the nonmetricity tensor $Q_{\alpha\mu\nu}$, and these are obtained by contracting the nonvanishing tensor $Q_{\alpha\mu\nu} \equiv \nabla_\alpha g_{\mu\nu}$.

In $f(Q,T)$ theories [42], the nonmetricity is coupled minimally to the trace of the matter energy–momentum tensor. The coupling between $Q$ and $T$ leads to nonconservation of the energy–momentum tensor, which has important physical implications such as changes to the thermodynamics of the Universe [43], the nongeodesic motion of test particles, and the appearance of an additional force.

In $f(R, L_m)$ theory [44], $f$ is a function of both the Ricci scalar and the matter Lagrangian. It is possible to have both an additive function such as

$$f(R, L_m) = \frac{R}{2} + L_m \tag{12}$$

or an exponential function such as

$$f(R, L_m) = \Lambda e^{\left[(\frac{1}{2\Lambda})R + (\frac{1}{\Lambda})(L_m)\right]}, \tag{13}$$

where $\Lambda > 0$ is an arbitrary constant. This function becomes

$$f(R, L_m) \approx \Lambda + \frac{R}{2} + L_m + ..., \tag{14}$$

in the limit $\left[(\frac{1}{2\Lambda})R + (\frac{1}{\Lambda})L_m\right] \ll 1$. The observed late time acceleration of the Universe can be described by $f(R, L_m)$ gravity [45, 46]. In [47], the MT wormhole solution was studied in $f(R, L_m)$ gravity assuming three different types of EOSs, namely linear barotropic, anisotropic, and isotropic EOSs. The wormhole solutions obey the flare-out condition for both the barotropic and the anisotropic cases. For the isotropic case, the shape function does not follow the flatness condition. The NEC is violated in the vicinity of the throat. The amount of exotic matter required near the wormhole throat is minimized in $f(R, L_m)$ gravity as compared to GR.

In the next two subsections, we take a deeper dive into $f(R)$ and Lovelock theories of gravity, which will set the required background to study wormholes in these theories. In Section 3, we provide a synopsis of the study of wormholes in each of these $f(R)$ theories of gravity.

## 2.2  f(R) Gravity Theories

We started the discussion of $f(R)$ theories in Section 2.1. To recap, in $f(R)$ theories [21, 22, 31, 48] $R$ is replaced by $f(R)$ as follows:

$$S = \int \frac{1}{2\kappa} f(R) \sqrt{-g}\, d^4x, \tag{15}$$

where $\kappa \equiv \frac{8\pi G}{c^4}$, and $c = 1$.

In $f(R)$ theories, there are two formalisms [20, 49] to derive the Einstein field equations from the action. They are (i) the metric formalism, in which a matter term $S_m(g_{\mu\nu}, \psi)$ is added to the action, where $\psi$ represents the matter field(s). The action is then varied with respect to the metric by not treating the connections $\Gamma^\mu_{\alpha\beta}$ independently to obtain the field equations. Then, there is (ii) the Palatini formalism, in which an independent variation is done with respect to the metric and the connection. The action is the same, but the curvature tensors and Ricci scalar are constructed with this independent connection.

Here, we will follow [21, 48] and show the derivation of the Einstein field equations in much more detail using the metric formalism. The connection coefficient $\Gamma^\alpha_{\beta\gamma}$ and the components of the Riemann curvature tensor $R^\alpha_{\beta\gamma\delta}$ are calculated using the standard equations

$$\Gamma^\alpha_{\beta\gamma} = \frac{1}{2} g^{\alpha\lambda}(g_{\lambda\beta,\gamma} + g_{\lambda\gamma,\beta} - g_{\beta\gamma,\lambda}), \tag{16}$$

and

$$R^\alpha_{\beta\gamma\delta} = \Gamma^\alpha_{\beta\delta,\gamma} - \Gamma^\alpha_{\beta\gamma,\delta} + \Gamma^\alpha_{\lambda\gamma}\Gamma^\lambda_{\beta\delta} - \Gamma^\alpha_{\lambda\delta}\Gamma^\lambda_{\beta\gamma}, \tag{17}$$

where the comma denotes partial derivatives. Before we vary the action, we first vary each of the quantities in the action. The variation of the determinant is

$$\delta\sqrt{-g} = -(1/2)\sqrt{-g}g_{\mu\nu}\delta g^{\mu\nu}. \tag{18}$$

The Ricci Scalar is $R = g^{\mu\nu}R_{\mu\nu}$, and the variation of $R$ with respect to $g^{\mu\nu}$ is

$$\begin{align}
\delta R &= R_{\mu\nu}\delta g^{\mu\nu} + g^{\mu\nu}\delta R_{\mu\nu}, \tag{19}\\
&= R_{\mu\nu}\delta g^{\mu\nu} + g^{\mu\nu}(\nabla_\rho \delta\Gamma^\rho_{\nu\mu} - \nabla_\nu \delta\Gamma^\rho_{\rho\mu}), \tag{20}
\end{align}$$

where

$$\delta\Gamma^\rho_{\mu\nu} = (1/2)g^{\rho\alpha}(\nabla_\mu \delta g_{\alpha\nu} + \nabla_\nu \delta g_{\alpha\mu} - \nabla_\alpha \delta g_{\mu\nu}), \tag{21}$$

$\Gamma^\rho_{\mu\nu}$ is the Christoffel symbol representing the Levi–Civita connection, and $\nabla_\mu$ is a covariant derivative. Now, by substituting (21) in (20), we obtain

$$\delta R = R_{\mu\nu}\delta g^{\mu\nu} + g_{\mu\nu}\Box\delta g^{\mu\nu} - \nabla_\mu\nabla_\nu\delta g^{\mu\nu}, \tag{22}$$

where $\Box \equiv g^{\alpha\beta}\nabla_\alpha\nabla_\beta$ is known as the D'Alembert operator. The variation of $f(R)$ is

$$\delta f(R) = \frac{df(R)}{dR}\delta R. \tag{23}$$

Let $f'(R) \equiv \frac{df(R)}{dR}$. Then,
$$\delta f(R) = f'(R)\delta R. \tag{24}$$

By varying the action (15), we obtain

$$\delta S = \int \frac{1}{2\kappa}\Big(\delta f(R)\sqrt{-g} + f(R)\delta\sqrt{-g}\Big)d^4x. \tag{25}$$

Substituting $\delta f(R)$ and $\delta\sqrt{-g}$ from (24) and (18) into (25), we obtain

$$\delta S = \int \frac{1}{2\kappa}\left(f'(R)\delta R\sqrt{-g} - \frac{1}{2}\sqrt{-g}g_{\mu\nu}\delta g^{\mu\nu}f(R)\right)d^4x. \tag{26}$$

Now, we substitute $\delta R$ from (22) into (26) to obtain

$$\delta S = \int \frac{\sqrt{-g}}{2\kappa} \left[ f'(R)\left(R_{\mu\nu}\delta g^{\mu\nu} + \left(g_{\mu\nu}\Box - \nabla_\mu\nabla_\nu\right)\delta g^{\mu\nu}\right) - \frac{1}{2}g_{\mu\nu}f(R)\delta g^{\mu\nu} \right] d^4x. \quad (27)$$

After integrating by parts and factoring out $\delta g^{\mu\nu}$, we obtain

$$\delta S = \int \frac{1}{2\kappa}\sqrt{-g}\delta g^{\mu\nu} \left[ f'(R)R_{\mu\nu} - \frac{1}{2}g_{\mu\nu}f(R) + \left(g_{\mu\nu}\Box - \nabla_\mu\nabla_\nu\right)f'(R) \right] d^4x. \quad (28)$$

Finally, by requiring that the action remain invariant with the variation of the metric, we obtain the field equations for $f(R)$ modified gravity theory:

$$f'(R)R_{\mu\nu} - \frac{1}{2}g_{\mu\nu}f(R) + \left(g_{\mu\nu}\Box - \nabla_\mu\nabla_\nu\right)f'(R) = \kappa T_{\mu\nu}, \quad (29)$$

where

$$T_{\mu\nu} = \frac{-2}{\sqrt{-g}}\frac{\delta(\sqrt{-g}\mathscr{L}_m)}{\delta g^{\mu\nu}}, \quad (30)$$

and $\mathscr{L}_m$ is the Lagrangian for matter.

## 2.3 Lovelock Gravity Theory

Introduced in 1971, Lovelock's theory of gravity [50] is considered to be the most generalized extension to the theory of gravitation in D dimensions, because it satisfies the requirements of GR that the field equations be covariant and not include more than the second-order derivatives of the metric.

We consider a generally covariant theory of gravity in D dimensions. The Lagrangian is a functional of independent variables $(g^{\mu\nu}, R^\mu{}_{\nu\rho\sigma})$, and the action is

$$S = \int_v d^D x \sqrt{-g} L[g^{\mu\nu}, R^\mu{}_{\nu\rho\sigma}] + S_m. \quad (31)$$

where $S_m$ is the action of the matter. Note that, depending on the purpose, different pairs of independent variables might be chosen to describe the system, such as $(g^{\mu\nu}, R_{\mu\nu\rho\sigma})$ or $(g^{\mu\nu}, R^{\mu\nu}{}_{\rho\sigma})$. However, the pair of $(g^{\mu\nu}, R^\mu{}_{\nu\rho\sigma})$ is the most appropriate one when deriving the field equations. For later convenience, we define the following:

$$P^{\mu\nu} := \left(\frac{\partial L}{\partial g_{\mu\nu}}\right)_{R_{\rho\sigma\lambda\gamma}}, \qquad P^{\mu\nu\rho\sigma} := \left(\frac{\partial L}{\partial R_{\mu\nu\rho\sigma}}\right)_{g_{\lambda\gamma}} \quad (32)$$

where, by definition, $P^{\mu\nu}$ is symmetric, and the *entropy tensor* $P^{\mu\nu\rho\sigma}$ follows the symmetries

$$P^{\mu\nu\rho\sigma} = -P^{\nu\mu\rho\sigma} = -P^{\mu\nu\sigma\rho}, \quad P^{\mu\nu\rho\sigma} = P^{\rho\sigma\mu\nu}, \quad P^{\mu[\nu\rho\sigma]} = 0. \quad (33)$$

We further construct the generalized form of the Ricci tensor as

$$\mathfrak{R}^{\mu\nu} = P^{\mu\rho\sigma\lambda}R^\nu{}_{\rho\sigma\lambda} \quad (34)$$

where it can be shown that [51, 52]

$$\left(\frac{\partial L}{\partial g^{\mu\nu}}\right)_{R^{\rho}{}_{\sigma\lambda\gamma}} = \mathfrak{R}_{\mu\nu} . \tag{35}$$

To obtain equations of motion, one needs to vary the action under the variation of the metric $g^{\mu\nu} \to g^{\mu\nu} + \delta g^{\mu\nu}$ that leads to the following variations

$$\delta g_{\mu\nu} = -g_{\mu\rho}g_{\nu\sigma} \,\delta g^{\rho\sigma} \tag{36}$$
$$\delta R^{\mu}{}_{\nu\rho\sigma} = g^{\mu\lambda}\Big(\nabla_{\rho}\nabla_{\nu}\delta g_{\sigma\lambda} - \nabla_{\rho}\nabla_{\lambda}\delta g_{\sigma\nu} - \rho \leftrightarrow \sigma\Big).$$

By varying the action (31), plugging in (36), and doing the partial integration, we obtain

$$\delta S = \int_v d^D x \sqrt{-g}\Big\{P_{\mu\nu} - \frac{1}{2}g_{\mu\nu}L + 2\nabla^{\rho}\nabla^{\sigma}P_{\mu\rho\sigma\nu}\Big\}\delta g^{\mu\nu} \tag{37}$$

where we have used the symmetry properties of $P_{\mu\nu\rho\sigma}$ as well. Finally, the field equations of motion read as $E_{\mu\nu} = (1/2)T_{\mu\nu}$, where

$$E_{\mu\nu} = \mathfrak{R}_{\mu\nu} - \frac{1}{2}g_{\mu\nu}L + 2\nabla^{\rho}\nabla^{\sigma}P_{\mu\rho\sigma\nu}, \tag{38}$$

and $T_{\mu\nu}$ is the energy–momentum tensor of matter. Moreover, $P_{\mu\nu}$ is replaced by the generalized Ricci tensor $\mathfrak{R}_{\mu\nu}$ via (35).

Given the fact that $P^{\mu\nu\rho\sigma}$ already contains second derivatives of the metric by definition, one realizes that the last term in the field equations (38) will involve up to fourth derivatives of $g_{\mu\nu}$. To ensure that we do not include derivatives higher than the second order, we impose the following condition:

$$\nabla_{\lambda}P^{\mu\nu\rho\sigma} = 0. \tag{39}$$

Even though one might think of imposing $\nabla_{\lambda}\nabla_{\gamma}P^{\mu\nu\rho\sigma} = 0$ as a more general condition, it turns out that this condition does not make any difference. Imposing (39) reduces the field equations to

$$\mathfrak{R}_{\mu\nu} - \frac{1}{2}g_{\mu\nu}L = \frac{1}{2}T_{\mu\nu} \tag{40}$$

which are nonlinear in second derivatives of the metric. In fact, imposing the linearity condition leads to the Einstein–Hilbert action. Here, one must be careful, since the form of the field equations is analogous to Einstein's field equations. However, $\mathfrak{R}_{\mu\nu}$ here is the generalized Ricci tensor defined in (34).

The next step is finding the form of the generalized Lagrangian that leads to field equations with no derivatives higher than the second order. This task reduces to finding those scalar functions of the metric and the Riemann tensor that satisfy (39). It can be shown that, at each order, such functions are uniquely constructed as

$$L_D^{(m)} = \frac{1}{16\pi}\frac{1}{2^m}\delta^{\mu_1\nu_1\cdots\mu_m\nu_m}_{\rho_1\sigma_1\cdots\rho_m\sigma_m}R^{\rho_1\sigma_1}{}_{\mu_1\nu_1}\cdots R^{\rho_m\sigma_m}{}_{\mu_m\nu_m}, \tag{41}$$

which determines the order m Lagrangian in D dimensions. It contains m factors of the Riemann tensor, and $\delta^{\mu_1\nu_1\cdots\mu_m\nu_m}_{\rho_1\sigma_1\cdots\rho_m\sigma_m}$ is a completely antisymmetric determinant tensor defined as

$$\delta^{\alpha\mu_1\nu_1\cdots\mu_m\nu_m}_{\beta\rho_1\sigma_1\cdots\rho_m\sigma_m} = \det \begin{bmatrix} \delta^\alpha_\beta & \delta^\alpha_{\rho_1} & \cdots & \delta^\alpha_{\sigma_m} \\ \delta^{\mu_1}_\beta & & & \\ \vdots & & \delta^{\mu_1\nu_1\cdots\mu_m\nu_m}_{\rho_1\sigma_1\cdots\rho_m\sigma_m} & \\ \delta^{\nu_m}_\beta & & & \end{bmatrix} . \qquad (42)$$

The entire Lanczos–Lovelock Lagrangian is then given by a sum over various orders of $L^{(m)}_D$, where

$$L_D = \sum_m c_m L^{(m)}_D. \qquad (43)$$

The expansion coefficients $c_m$ are initially arbitrary. The detailed explanation of the construction procedure of Lanczos–Lovelock (LL) Lagrangian and the proof of its uniqueness are out of the scope for this review. So, we refer the interested reader to the original papers by Lanczos [53, 54] and Lovelock [50].

Note that the order m Lagrangian in (41) has been written in terms of $R^{\mu\nu}{}_{\rho\sigma}$. This change of variable enables us to use the following identity that simplifies our guesses for a generalized form of the Lagrangian, where

$$\left(\frac{\partial L}{\partial g^{\mu\nu}}\right)_{R^{\rho\sigma}{}_{\lambda\gamma}} = 0. \qquad (44)$$

This identity implies that the Lagrangian must be independent of the metric tensor, and hence, indices of the Riemann tensor $R^{\mu\nu}{}_{\rho\sigma}$ must be contracted only by the Kronecker deltas in the form that we have in (41).

The zeroth order Lagrangian has $m = 0$, which simply leads to a constant term (e.g., cosmological constant). In the first order, we obtain

$$L^{(1)} = \frac{1}{32\pi} \delta^{\mu\rho}_{\nu\sigma} R^{\nu\sigma}{}_{\mu\rho} = R . \qquad (45)$$

So, the first order of the LL Lagrangian gives back the Einstein–Hilbert Lagrangian. As expected, the generalized equations of motion (40) simply reduce to the Einstein fields equations for $L^{(0)} + L^{(1)}$, i.e.,

$$\Lambda g_{\mu\nu} + G_{\mu\nu} = \kappa T_{\mu\nu}, \qquad (46)$$

where $\Lambda$ is the cosmological constant, and $G_{\mu\nu} = R_{\mu\nu} - \frac{1}{2} R g_{\mu\nu}$ is the well-known Einstein tensor.

To study the wormhole solutions of generalized gravity discussed in [18, 55], we go up to the third order in the LL Lagrangian:

$$S = \int_v d^D x \sqrt{-g} \left( \alpha_1 L^{(1)} + \alpha_2 L^{(2)} + \alpha_3 L^{(3)} \right) + S_m \qquad (47)$$

where the next two orders are determined as

$$L^{(2)} = \frac{1}{16\pi}\Big(R_{\mu\nu\gamma\delta}R^{\mu\nu\gamma\delta} - 4R_{\mu\nu}R^{\mu\nu} + R^2\Big), \qquad (48)$$

$$\begin{aligned}L^{(3)} &= \frac{1}{16\pi}\Big(2R^{\mu\nu\sigma\kappa}R_{\mu\kappa\rho\tau}R^{\rho\tau}{}_{\mu\nu} + 8R^{\mu\nu}{}_{\sigma\rho}R^{\sigma\kappa}{}_{\nu\tau}R^{\rho\tau}{}_{\mu\kappa} + 24R^{\mu\nu\sigma\kappa}R_{\sigma\kappa\nu\rho}R^{\rho}{}_{\mu} + 3RR^{\mu\nu\sigma\kappa}R_{\sigma\kappa\mu\nu}\\ &\quad + 24R^{\mu\nu\sigma\kappa}R_{\sigma\mu}R_{\kappa\nu} + 16R^{\mu\nu}R_{\nu\sigma}R^{\sigma}{}_{\mu} - 12RR^{\mu\nu}R_{\mu\nu} + R^3\Big),\end{aligned} \qquad (49)$$

and $\alpha_m$ are the Lovelock coefficients. The second order term (48) is known as the Gauss–Bonnet Lagrangian.

The generalized field equation (40) corresponding to various orders of the Lagrangian can be written separately as well,

$$\Lambda g_{\mu\nu} + G_{\mu\nu} + \sum_{m\geq 2}\alpha_m\big(\mathcal{R}^{(m)}_{\mu\nu} - \frac{1}{2}g_{\mu\nu}L^{(m)}\big) = \kappa_m T_{\mu\nu}, \qquad (50)$$

and $T_{\mu\nu}$ is the energy–momentum tensor.

The field equations corresponding to the second and third order Lagrangian are respectively determined by $\mathfrak{R}^{(2)}_{\mu\nu}$ and $\mathfrak{R}^{(3)}_{\mu\nu}$ which are

$$\mathfrak{R}^{(2)}_{\mu\nu} \equiv \frac{1}{8\pi}\Big(R_{\mu\sigma\kappa\tau}R_{\nu}{}^{\sigma\kappa\tau} - 2R_{\mu\rho\nu\sigma}R^{\rho\sigma} - 2R_{\mu\sigma}R^{\sigma}{}_{\nu} + RR_{\mu\nu}\Big), \qquad (51)$$

and

$$\begin{aligned}\mathfrak{R}^{(3)}_{\mu\nu} &\equiv \frac{-3}{16\pi}\Big(4R^{\tau\rho\sigma\kappa}R_{\sigma\kappa\lambda\rho}R^{\lambda}{}_{\nu\tau\mu} - 8R^{\tau\rho}{}_{\lambda\sigma}R^{\sigma\kappa}{}_{\tau\mu}R^{\lambda}{}_{\nu\rho\kappa} + 2R_{\nu}{}^{\tau\sigma\kappa}R_{\sigma\kappa\lambda\rho}R^{\lambda\rho}{}_{\tau\mu}\\ &\quad - R^{\tau\rho\sigma\kappa}R_{\sigma\kappa\tau\rho}R_{\nu\mu} + 8R^{\tau}{}_{\nu\sigma\rho}R^{\sigma\kappa}{}_{\tau\mu}R^{\rho}{}_{\kappa} + 8R^{\sigma}{}_{\nu\tau\kappa}R^{\tau\rho}{}_{\sigma\mu}R^{\kappa}{}_{\rho}\\ &\quad + 4R_{\nu}{}^{\tau\sigma\kappa}R_{\sigma\kappa\mu\rho}R^{\rho}{}_{\tau} - 4R_{\nu}{}^{\tau\sigma\kappa}R_{\sigma\kappa\tau\rho}R^{\rho}{}_{\mu} + 4R^{\tau\rho\sigma\kappa}R_{\sigma\kappa\tau\mu}R_{\nu\rho} + 2RR_{\nu}{}^{\kappa\tau\rho}R_{\tau\rho\kappa\mu}\\ &\quad + 8R^{\tau}{}_{\nu\mu\rho}R^{\rho}{}_{\sigma}R^{\sigma}{}_{\tau} - 8R^{\sigma}{}_{\nu\tau\rho}R^{\tau}{}_{\sigma}R^{\rho}{}_{\mu} - 8R^{\tau\rho}{}_{\sigma\mu}R^{\sigma}{}_{\tau}R_{\nu\rho} - 4RR^{\tau}{}_{\nu\mu\rho}R^{\rho}{}_{\tau}\\ &\quad + 4R^{\tau\rho}R_{\rho\tau}R_{\nu\mu} - 8R^{\tau}{}_{\nu}R_{\tau\rho}R^{\rho}{}_{\mu} + 4RR_{\nu\rho}R^{\rho}{}_{\mu} - R^2 R_{\mu\nu}\Big).\end{aligned} \qquad (52)$$

# 3 Fundamentals of Morris-Thorne (MT) Wormhole

In Section 3.1, we give a brief background of how the stability of a wormhole (usually an MT wormhole) is studied in GR. In Section 3.2, we summarize a general methodology to study wormholes in $f(R)$ gravity theories based on the various papers we reviewed on this subject. This background will be useful to follow Section 4, where we review these calculations in greater detail for general $f(R)$ theories and Lovelock gravity theory.

## 3.1 MT Wormhole Stabilization in GR

In [9], they explain in great detail why blackholes and Schwarzschild wormholes are not traversable. However, the MT wormhole is designed to be made traversable if it has the following properties that ensure wormhole stability for traversability. The MT metric allows for the realization of faster-than-light interstellar space travel that does not violate the special relativistic light speed limit. The metric should be spherically symmetric and static (time independent). The following metric has these properties [9]:

$$ds^2 = -e^{2\Phi}dt^2 + [1 - (b/r)]^{-1}dr^2 + r^2[d\theta^2 + \sin^2\theta d\phi^2]. \tag{53}$$

The solution must obey the Einstein field equations as does (53). The solution must have a throat that connects two asymptotically flat regions of spacetime. The spatial geometry must have a wormhole shape consistent with the well-known Flamm diagram for a spherically symmetric throat. This puts the following constraints on the shape function $b(r)$ and redshift function $\Phi(r)$:

- The throat is at minimum of r, which is specified as $r_0$.
- $b(r)$ is finite, continuous, and differentiable.
- In this spacetime, $(1 - b/r) \geq 0$, which implies $b/r \leq 1$, and so $b(r) \leq r$.
- Proper radial distance is defined by

$$l \equiv \int_{r_0}^{r} \frac{1}{\sqrt{1 - (b/r)}} dr,$$

and should be real and finite for $r > r_0$.
- As $l \to \pm\infty$ (asymptotically flat regions of the Universe), $b/r \to 0$, and so $r \sim |l|$.
- There should be no horizons, since it will prevent two-way travel through the wormhole. There are no singularities. This implies that $\Phi$ is finite, continuous, and differentiable everywhere, as well as the fact that $\tau$ measuring proper time in asymptotically flat regions implies $\Phi \to 0$ as $l \to \pm\infty$.
- The flare-out condition is

$$\frac{(b - b'r)}{2b^2} > 0,$$

and so $rb' < b$. That is, the throat of the wormhole must expand outward from the central point. The throat of the wormhole must open up as one travels through it.

The tidal gravitational forces experienced by a traveler must be $\leq g$, where g is the acceleration due to Earth's gravity. This condition is not a rigid requirement.

The procedure to check the stability of this traversable wormhole in GR involves the following steps:

**Compute Curvature tensors**: Here, we give as briefly as possible the method to calculate the curvature tensors. First, using the MT metric written in the form

$$ds^2 = g_{\alpha\beta}dx^\alpha dx^\beta, \tag{54}$$

with $x^0 = t$, $x^1 = r$, $x^2 = \theta$, and $x^3 = \phi$, the connection coefficient $\Gamma^\alpha_{\beta\gamma}$ and the components of the Riemann curvature tensor $R^\alpha_{\beta\gamma\delta}$ are calculated using the standard Equations (16) and (17). By applying these equations to the metric, we can obtain the 24 nonzero components of the Riemann tensor, as shown in Equation (5) of [9]. These were obtained using the basis vectors $(e_t, e_r, e_\theta, e_\phi)$.

To further simplify the calculations, we can switch to the following orthonormal basis vectors:

- $\hat{e}_t = e^{-\Phi} e_t,$
- $\hat{e}_r = (1 - \frac{b}{r})^{-1/2} e_r;$
- $\hat{e}_\theta = r^{-1} e_\theta;$
- $\hat{e}_\phi = (r \sin\theta)^{-1} e_\phi.$

In this basis, the metric coefficients become the same as those of flat (Minkowski) spacetime,

$$g_{\alpha\beta} = \hat{e}_\alpha \hat{e}_\beta = \eta_{\alpha\beta} = \begin{bmatrix} -1 & 0 & 0 & 0 \\ 0 & 1 & 0 & 0 \\ 0 & 0 & 1 & 0 \\ 0 & 0 & 0 & 1 \end{bmatrix}, \tag{55}$$

and the 24 nonzero components of the Riemann tensor take a much simpler form, as seen in Equation (8) of [9].

**Contraction**: Next, by contracting the Riemann tensor, we obtain the Ricci tensor,

$$R_{\mu\nu} = R^\alpha_{\mu\alpha\nu}, \tag{56}$$

and again, by contracting the Ricci tensor, we obtain the Ricci scalar R.

**Compute Einstein tensor**: We finally obtain the Einstein tensor $G_{\mu\nu}$ from the metric, Ricci tensor, and the Ricci scalar. This forms the left-hand side of the Einstein field equations, as given in (46). This yields the nonzero components of $G_{\mu\nu}$ in terms of the shape function and redshift function, namely

$$G_{tt} = \frac{b'}{r^2}, \tag{57}$$

$$G_{rr} = \frac{-b}{r^3} + 2(1 - \frac{b}{r})\frac{\Phi'}{r}, \tag{58}$$

and

$$G_{\theta\theta} = G_{\phi\phi} = (1 - \frac{b}{r})\left[\Phi'' - \frac{(b'r - b)}{2r(r - b)}\Phi' + (\Phi')^2 + \frac{\Phi'}{r} - \frac{(b'r - b)}{2r^2(r - b)}\right]. \tag{59}$$

**Compute stress–energy tensor**: The next step involves computing the stress–energy tensor (the right-hand side of the Einstein field equations). Based on Birkhoff's theorem, which states that "any spherically symmetric solution of the vacuum field equations must be static and asymptotically flat", the exterior solution must be given by the Schwarzschild metric (with certain modifications), which is a spherical wormhole. Therefore, we cannot have a vacuum solution for a traversable wormhole, which implies that our wormhole must be threaded by matter with a nonzero stress–energy tensor. Based on Einstein's field equations,

$$G_{\mu\nu} = \kappa T_{\mu\nu}. \tag{60}$$

$T_{\mu\nu}$ must have the same algebraic structure as $G_{\mu\nu}$ in the orthonormal basis that we have chosen. Similar to $G_{\mu\nu}$, only the four components $T_{tt}$, $T_{rr}$, $T_{\theta\theta}$, and $T_{\phi\phi}$ are nonzero. Based on a remote static observer's measurement, each of the components have a simple physical interpretation as follows:

$$T_{tt} = \rho(r) \tag{61}$$

is the total rest energy density that the static observer measures,

$$T_{rr} = -\tau(r) \tag{62}$$

is the radial tension measured per unit area, and

$$T_{\theta\theta} = T_{\phi\phi} = p(r) \tag{63}$$

is the tangential pressure measured per unit area in a direction orthogonal to the radial tension $\tau(r)$. This give us the final stress–energy tensor:

$$T_{\mu\nu} = \begin{bmatrix} \rho(r) & 0 & 0 & 0 \\ 0 & -\tau(r) & 0 & 0 \\ 0 & 0 & p(r) & 0 \\ 0 & 0 & 0 & p(r) \end{bmatrix}. \tag{64}$$

**Engineer the traversable wormhole**: In the next step, we need to 'engineer' the traversable wormhole to obtain the properties enumerated earlier in this section. This can be done by controlling the shape function $b(r)$ and the redshift function $\Phi(r)$ based on a suitable $T_{\mu\nu}$ that we require. We substitute the $G_{\mu\nu}$ and $T_{\mu\nu}$ found in the previous steps into (60) to solve for $\rho(r)$, $\tau(r)$, and $p(r)$ in terms of $b(r)$ and $\Phi(r)$ to obtain

$$\rho(r) = \frac{b'}{r^2}, \tag{65}$$

$$\tau(r) = \frac{b}{r^3} - \frac{2(r-b)}{r^2}\Phi', \tag{66}$$

$$p(r) = \frac{r}{2}\left[(\rho - \tau)\Phi' - \tau'\right] - \tau. \tag{67}$$

Our strategy for the stabilization of the MT wormhole in Einstein gravity will involve tailoring $b(r)$ and $\Phi(r)$ to build a wormhole with the required properties. Our choice of $b(r)$ gives us $\rho(r)$. Our choice of $b(r)$ and $\Phi(r)$ will give us the tangential pressure $\tau(r)$. We next find $p(r)$ using $\rho(r)$, $\tau(r)$, and $\Phi(r)$. We then ensure that the averaged NEC and WEC are violated, which means that the source of matter for traversable wormholes must be exotic.

With this process for analyzing the stability of the MT wormhole in GR, many authors in our review have used similar techniques to analyze the stability of the Morris–Thorne wormhole in $f(R)$ gravity theories. We will hone into the nuances of these analysis in the following sections.

## 3.2 A General Methodology for MT Wormhole Stabilization in f(R) Gravity Theories

Based on Section 4, we briefly summarize the method that can be used to stabilize the MT wormhole in $f(R)$ gravity theories. First, we require that matter threading the wormhole satisfy the NEC and WEC. We include the required flare-out condition for traversable wormholes. We delegate the required energy condition violations and sustenance of the non-standard wormhole geometry to the higher-order curvature terms, including the derivative terms ($T_{\mu\nu}^{(c)}$). Consider a redshift function ($\Phi = c$, $\Phi' = 0$), where $c$ is a constant, which simplifies calculations and still provides physically relevant solutions. This condition defines an ultrastatic traversable wormhole and is a very narrow and specific subclass of solutions called zero tidal force (ZTF) solutions. We specify $b(r)$. Some examples used in [48] are $b(r) = \frac{r_0^2}{r}$, $b(r) = \sqrt{r_0 r}$, and $b(r) = r_0 + \gamma^2 r_0(1 - \frac{r_0}{r})$, with $0 < r < 1$. For each shape function $b(r)$, we assume an equation of state such as $p_r = p_r(\rho)$ or $p_t = p_t(\rho)$. We find $F(r)$ (see Section 4.1) from the modified gravitational field equations, with the Ricci curvature scalar R(r) obtained from the MT metric. Finally, we obtain the exact $f(R)$ that we need from the trace Equation (68).

# 4 Wormholes in f(R) Gravity Theories

## 4.1 Wormholes in f(R) Gravity Theory

In this section, we mainly follow the references [21, 22, 31, 48] to analyze the stability of an MT wormhole in $f(R)$ gravity theories. In Section 2.2, we obtained (29) for $f(R)$ gravity theory. Contracting this field equation, we obtain the trace of $T_{\mu\nu}$,

$$FR - 2f(R) + 3\Box F = T, \tag{68}$$

where $F = f'(R)$ is used for convenience, R is the Ricci scalar, and $T = T^\mu_\mu$ is the trace of the stress–energy tensor $T_{\mu\nu}$. By substituting (68) into (29) and the rearranging terms, we obtain an updated field equation:

$$G_{\mu\nu} \equiv R_{\mu\nu} - \frac{1}{2}R g_{\mu\nu} = T^{\text{eff}}_{\mu\nu}. \tag{69}$$

We will call (69) the effective field equation for $f(R)$ gravity theory, where

$$T^{\text{eff}}_{\mu\nu} = T^{(c)}_{\mu\nu} + T^{(m)}_{\mu\nu}. \tag{70}$$

$T^{(c)}_{\mu\nu}$ is the curvature stress–energy tensor for a higher-order curvature given by

$$T^{(c)}_{\mu\nu} = \frac{1}{F}\left[\nabla_\mu \nabla_\nu F - \frac{1}{4}g_{\mu\nu}\left(RF + \Box F + T\right)\right], \tag{71}$$

and $T^{(m)}_{\mu\nu} = \frac{T_{\mu\nu}}{F}$. Here, we write the energy–momentum tensor in terms of an anisotropic distribution of matter as follows:

$$T_{\mu\nu} = (\rho + p_t)u_\mu u_\nu + p_t g_{\mu\nu} + (p_r - p_t)x_\mu x_\nu, \tag{72}$$

such that $u^\mu u_\mu = -1$ and $x^\mu x_\mu = -1$, where $u^\mu$ is a four-velocity vector,

$$x^\mu = \sqrt{1 - \frac{b'(r)}{r}}\delta^\mu_r$$

is a unit spacelike vector in the radial direction, $\rho(r)$ is the rest energy density, $p_r(r)$ is the radial tension, and $p_t(r)$ is the tangential pressure orthogonal to $x^\mu$. This gives us

$$T_{\mu\nu} = \begin{bmatrix} -\rho(r) & 0 & 0 & 0 \\ 0 & p_r(r) & 0 & 0 \\ 0 & 0 & p_r(r) & 0 \\ 0 & 0 & 0 & p_t(r) \end{bmatrix} \tag{73}$$

and
$$T = (T^\mu_\mu) = (-\rho + 2p_r + p_t). \tag{74}$$

By substituting this value of T in the trace form of the field equations for $f(R)$ gravity (68), we obtain
$$FR - 2f + 3\Box F = (-\rho + 2p_r + p_t). \tag{75}$$

We now use the MT wormhole metric (53) just as we did in GR, where $\Phi(r)$ is the redshift function, and $b(r)$ is the shape function as before. The radial coordinate r decreases from $\infty$ to a minimum value $r_0$ at the wormhole throat. At the throat, $b(r_0) = r_0$, and it then increases from $r_0$ back to $\infty$. We recall the flare-out condition, which is important for traversability:
$$\frac{(b - b'(r))}{b^2} > 0. \tag{76}$$

At the throat, $b(r_0) = r = r_0$ $b(r_0) > b'(r_0)r$ or $r > b'(r_0)r$, $b'(r_0) < 1$ is the flare-out condition that we need for a traversable wormhole solution. For the wormhole to be traversable, no horizon should be present. Horizons are surfaces with $e^{2\Phi} \to 0$, so we want $\Phi(r)$ to be finite everywhere. This implies that we can use the constant redshift functions $\Phi(r)$, $\Phi = c$, and $\Phi' = 0$. This will help to simplify calculations and avoid fourth-order differential equations. The effect of using a variable redshift function is discussed in Section 4.3. The following steps can be used to calculate $\rho$, $p_r$, and $p_t$ in terms of the shape and redshift functions. Substitute $T_{\mu\nu}$ in the effective field Equation (69). Use the MT metric (53) to obtain $R = \frac{2b'}{r}$. Use
$$\Box F = (1 - b/r) \left[ F'' - \frac{b'r - b}{2r^2(1 - b/r)} F' + \frac{2F'}{r} \right]. \tag{77}$$

Define
$$H(r) \equiv \frac{1}{4}(FR + \Box F + T). \tag{78}$$

Note that $F' = \frac{d}{dR}(\frac{df(R)}{dR})$, and $F'' = \frac{d(F'(R))}{dR}$. Thus, we obtain the following simplified equations from the effective field Equation (69):
$$\rho = \frac{Fb'}{r^2}, \tag{79}$$

$$p_r = \frac{-bF}{r^3} + \frac{F'}{2r^2}(b'r - b) - F''(1 - b/r), \tag{80}$$

and
$$p_t = -\frac{F'}{r}(1 - b/r) + \frac{F}{2r^3(b - b'r)}. \tag{81}$$

These are the required simplified equations for matter threading the wormhole as a function of $b(r)$ and $F(r)$. We will use them later in this paper. We can now determine the matter content by choosing an appropriate shape function and a specific form of $F(r)$. At this point, let us recap the strategy for stabilizing a wormhole in $f(R)$ gravity. We choose a shape function $b(r)$. Then, we specify an equation of state such as $p_r = p_r(\rho)$ or $p_t = p_t(\rho)$. This will let us compute $F(r)$ from the effective field equation of Equation (69). We can also find the Ricci scalar R(r) from the MT wormhole metric (53). Then, we obtain $T = T^\mu_\mu$ as a function of r. Finally, we compute the function $f(R)$ from the trace of the field Equation (68).

## 4.2 Violation of Energy Conditions

Just as the motion of a single particle is governed by the geodesic equation, the equation of motion of a family of particles, also known as a congruence, is governed by the Raychaudhuri equation [56]. From this equation, the following focusing condition in terms of the Ricci tensor arises:

$$R_{\mu\nu}k^\mu k^\nu \geq 0,$$

where $k^\mu$ is a null vector. If this condition is satisfied, then the geodesic congruences focus into a finite value of the parameter for labeling points on the geodesics. In GR, this is written as the NEC, in terms of the stress–energy tensor, as follows:

$$T_{\mu\nu}k^\mu k^\nu \geq 0.$$

In modified gravity, in particular in $f(R)$ theories, we could first assume that $T^{(m)}_{\mu\nu}$ satisfies this energy condition, and violation of the energy condition can be assumed to come from the higher-order curvature terms $T^{(c)}_{\mu\nu}$. Note that this condition applied to $f(R)$ gravity theory does not mean geodesics are focused, as required by the Raychaudhuri equation. In terms of the radial null vector, violation of the NEC requires

$$T^{\text{eff}}_{\mu\nu}k^\mu k^\nu < 0, \tag{82}$$

and takes the form

$$\rho^{\text{eff}} + p_r^{\text{eff}} = \frac{\rho + p_r}{F} = \frac{1}{F}(1 - b/r)\left[F'' - F'\frac{b'r - b}{2r^2(1 - b/r)}\right], \tag{83}$$

where

$$\rho^{\text{eff}} + p_r^{\text{eff}} < 0.$$

Using the gravitational field equation of Equation (69), we obtain

$$\rho^{\text{eff}} + p_r^{\text{eff}} = \frac{b'r - b}{r^3}. \tag{84}$$

By applying the flare-out condition,

$$\frac{b'r - b}{b^2} < 0,$$

$$\rho^{\text{eff}} + p_r^{\text{eff}} < 0.$$

At the throat, with $r = r_0$,

$$\rho^{\text{eff}} + p_r^{\text{eff}}|_{r_0} = \frac{\rho + p_r}{F}\bigg|_{r_0} + \frac{1 - b'(r_0)}{2r_0}\frac{F'}{F}\bigg|_{r_0} < 0. \tag{85}$$

At the throat, this gives us

$$F'_0 < -2r_0 \frac{(\rho + p_r)}{(1-b')}|_{r_0} ; \qquad F > 0 \tag{86}$$

$$F'_0 > -2r_0 \frac{(\rho + p_r)}{(1-b')}|_{r_0} ; \qquad F < 0 \tag{87}$$

We have been highlighting that matter threading the wormhole should obey the WEC as well, i.e., $\rho \geq 0$, and $\rho + p_r \geq 0$. Applying these WECs to the simplified Equations (79)–(81) for $\rho$, $p_r$, and $p_t$, we obtain

$$\frac{Fb'}{r^2} \geq 0 \tag{88}$$

$$\frac{(2F + rF')(b'r - b)}{2r^2} - F''(1 - b/r) \geq 0. \tag{89}$$

The four inequalities above (86)–(89) must be obeyed by the function $f(R)$ for traversable wormholes in $f(R)$ theories, thus taking into account that matter threading the wormhole satisfies the NEC and WEC, and the flare-out conditions are satisfied at the wormhole throat. The task of maintaining the wormhole geometry (violating the NEC) is delegated to the higher-order curvature terms $T^{(c)}_{\mu\nu}$.

## 4.3 Effect of Using a Variable Redshift Function

The effect of using a variable redshift function for wormholes has been studied by [57] in $\kappa(R,T)$ gravity with $\Phi(r) = \frac{\alpha}{r}$, where $\alpha$ is a constant. The authors of [57] obtained solutions that require exotic matter. Wormholes in $f(R)$ gravity were studied by [58–62] with a variable redshift function $\Phi(r) = \ln\left(\frac{r_0}{r} + 1\right)$, and the authors in [58–62] found solutions that require nonexotic matter. The authors in [63] studied the effect of a variable redshift function $\Phi(r) = -\frac{\alpha}{r}$ with $\alpha \geq 0$, where $\alpha$ is a constant, and [64] used $\Phi(r) = \frac{\alpha}{r}$. [21] in particular studied wormholes in f(R) gravity with a variable redshift function $\Phi(r) = \frac{1}{r}$ in great detail. Here, we discuss the solution obtained by them. [21] used the function $f(R) = R - \mu R_c (\frac{R}{R_c})^p$, where $\mu$, $R_c$, and $p$ are constants, with $\mu > 0$, $R_c > 0$, and $0 < p < 1$. [21] also used the shape function $b(r) = \frac{r}{e^{r-r_0}}$. Their motivation for selecting the above $f(R)$ function is related to obtaining viable dark energy models, and it is also related to finding explanations for the accelerated expansion of the universe. Based on past studies, this $f(R)$ function seems to be a good candidate for exploring wormholes in $f(R)$ gravity with both a constant and a variable redshift function.

With the redshift function $\Phi = \frac{1}{r}$, they found $\rho > 0$ for $r \geq 1.2$, $\rho + p_r > 0$ for $r \geq 1.2$, $\rho + p_t > 0$ for $r > 1.8$, $\rho + p_r + 2p_t < 0$ for $r > 1.2$, $\rho - |p_r| > 0$ for $r \geq 1.2$, and $\rho - |p_t| > 0$ for $r > 1.8$, and the NEC, WEC, and DEC were satisfied for $r > 1.8$ with the variable redshift function. So in conclusion, for wormholes with $r_0 > 1.8$—where $r_0$ is the radius of the throat—and with variable redshift function $\frac{1}{r}$, they obtained wormhole geometries free of exotic matter. They also did similar calculations in GR with the variable redshift function $\frac{1}{r}$ and found that there was no solution in GR without exotic matter for any value of r.

## 4.4 The Stability Condition and the Speed of Sound

In [65], the authors analyzed the stability of a wormhole in addition to traversability in $f(R)$ gravity. The authors in [65] used a power law $f(R) = f_0 R^{1+\epsilon}$, which can be written in the form $f(R) = R + \epsilon R ln(R) + O(\epsilon^2)$, for $\epsilon \ll 1$, to consider small deviations from Einstein gravity. Such an approach was used in the study of compact objects such as neutron stars and blackholes [66, 67], where departure from GR can be useful to explain observations.

For the analysis, the authors used the MT wormhole metric with a redshift function $2\Phi(r) = \frac{r_0}{r}$ and two different shape functions $\frac{b(r)}{r} = \left(\frac{r_0}{r}\right)^{1+\beta}$, where $\beta$ is a real number, $\beta + 1 > 0$, and $\frac{b(r)}{r} = \frac{r_0}{1+\alpha r}$—with $\alpha \equiv \frac{r_0 - 1}{r_0}$—to ensure that the wormhole is not singular at $r_0$. They obtained the lengthy Equations (13)–(16) in [65] for $\rho(r)$, $p_r(r)$, $p_t(r)$, and the average pressure $p(r) = \frac{1}{3}[p_r(r) + 2p_t(r)]$, respectively. Next, they used ideas from fluid dynamics [68–70] and defined the adiabatic speed of sound:

$$c_s^2 = \left(\frac{\partial p}{\partial \rho}\right)_s. \tag{90}$$

To obtain a stability condition, we need a vanishing speed of sound at the throat:

$$\frac{dp}{d\rho}\bigg|_{r_0} = 0 \tag{91}$$

By inserting Equations (13) and (14) in [65] for $\rho(r)$ and $p_r(r)$ into Equation (91) above, they obtained the required stability conditions, namely Equations (24) and (25) in [65], for $\frac{b(r)}{r} = \left(\frac{r_0}{r}\right)^{\beta+1}$ and for $\frac{b(r)}{r} = \frac{r_0}{1+\alpha r}$, respectively. Their primary finding was that for a specific range of values of $\epsilon$ in the assumed $f(R)$, and using the condition for vanishing speed of sound, stable wormhole solutions exist in the presence of nonexotic matter (a perfect fluid).

## 4.5 Wormholes in Lovelock Gravity Theory

We will now discuss how the modifications in Lovelock gravity help with the stabilization of a wormhole. In this section, we mainly follow [18, 55]. The second- and third-order curvature terms in Lovelock gravity are important to be considered, especially in the case of wormholes with smaller throat diameters, where the curvature is very high. First, the MT wormhole metric is modified for n dimensions as follows:

$$ds^2 = -e^{2\Phi(r)}dt^2 + \left(1 - b(r)/r\right)^{-1} dr^2 + r^2 \left(d\theta_1^2 + \sum_{i=2}^{n-2} \prod_{j=1}^{i-1} \sin^2\theta_j \, d\theta_i^2\right), \tag{92}$$

where $\Phi(r)$ is the redshift function, and $b(r)$ is the shape function. The WEC requires that matter threading the wormhole have positive energy density $\rho(r)$ and positive $(\rho(r) - \tau(r))$, where $\tau(r)$ is the radial tension, as well as positive $\rho(r) + p(r)$, where $p(r)$ is the tangential pressure orthogonal to the radius. If the WEC

$$\rho = T_{\mu\nu}u^\mu u^\nu \geq 0$$

is satisfied everywhere, then the wormhole can be constructed with normal matter without the need for exotic matter. Here, $u_\mu$ is a timelike velocity of the observer. Note that the stress tensor $T_{\mu\nu}$ used in the WEC is calculated using (50), which includes all the Lovelock gravity modifications to Einstein's field equations.

The NEC is
$$T_{\mu\nu} k^\mu k^\nu \geq 0$$

The authors in [18] used the orthonormal basis set
$$\mathbf{e}_{\hat{t}} = e^{-\Phi} \frac{\partial}{\partial t}, \tag{93}$$

$$\mathbf{e}_{\hat{r}} = \left(1 - \frac{b(r)}{r}\right)^{1/2} \frac{\partial}{\partial r}, \tag{94}$$

$$\mathbf{e}_{\hat{1}} = r^{-1} \frac{\partial}{\partial \theta_1}, \tag{95}$$

and
$$\mathbf{e}_{\hat{i}} = \left(r \prod_{j=1}^{i-1} \sin \theta_j\right)^{-1} \frac{\partial}{\partial \theta_i} \tag{96}$$

to calculate the components of the energy–momentum tensor, which gives us $\rho$, $\tau$, and $p$. That is,
$$T_{tt} = \rho, \tag{97}$$
$$T_{rr} = -\tau, \tag{98}$$

and
$$T_{ii} = p. \tag{99}$$

Then, by setting the $\Phi(r) = c = 0$, where $c$ is a constant, $(\rho - \tau)$ and $(\rho + p)$ are calculated as follows:
$$(\rho - \tau) = -\frac{(n-2)}{2r^3}(b - rb')\left(1 + \frac{2\alpha_2 b}{r^3} + \frac{3\alpha_3 b^2}{r^6}\right) \tag{100}$$

and
$$(\rho + p) = -\frac{(b - rb')}{2r^3}\left(1 + \frac{6\alpha_2 b}{r^3} + \frac{15\alpha_3 b^2}{r^6}\right) + \frac{b}{r^3}\left\{(n-3) + (n-5)\frac{2\alpha_2 b}{r^3} + (n-7)\frac{3\alpha_3 b^2}{r^6}\right\}, \tag{101}$$

where $\alpha_2$ and $\alpha_3$ are the second-order and third-order Lovelock coefficients. In order to obtain positive $\rho$ and $(\rho + p)$ values, the rest of the analysis was done by studying three types of shape functions $b(r)$: the power law, the logarithmic, and the hyperbolic shape functions. The power law shape function is given by
$$b = \frac{r_0^m}{r^{m-1}}, \tag{102}$$

with positive m. The functions $\rho$ and $(\rho + p)$ for the power law above are positive for $r > r_0$, provided that $r_0 > r_c$, and $r_c$ is the largest positive real root of certain equations given in [18]. For the logarithmic shape function, we have
$$b(r) = \frac{r \ln(r_0)}{\ln(r)}, \tag{103}$$

where $\rho$ and $(\rho + p)$ are positive for $r > r_0$, with the condition $r_0 \geq r_c$, and $r_c$ is the largest real root of a second set of equations in [18]. For the hyperbolic shape function, we have

$$b(r) = \frac{r_0 \tanh(r)}{\tanh(r_0)}, \tag{104}$$

where $\rho$ and $(\rho + p)$ are positive if $r_0 > r_c$, with the condition that $r_c$ is the largest root of a third set of equations in [18].

To ensure that $(\rho - \tau)$ is positive, we need

$$1 + \frac{2\alpha_2 b}{r^3} + \frac{3\alpha_3 b^2}{r^6} < 0,$$

where $\alpha_2$ and $\alpha_3$ are the Lovelock coefficients. For certain combinations of the Lovelock coefficients contributing to the throat radius, with a negative value for either $\alpha_2$ or $\alpha_3$, the above condition is satisfied only in the vicinity of the throat for all three types of shape functions discussed above. The throat radius must be in the range $r_- < r_0 < r_+$ with

$$r_- = \sqrt{-\alpha_2 - \sqrt{\alpha_2^2 - 3\alpha_3}}, \tag{105}$$

and

$$r_+ = \sqrt{-\alpha_2 + \sqrt{\alpha_2^2 - 3\alpha_3}}. \tag{106}$$

The following points summarize why the higher-order curvature terms are needed to stabilize a wormhole, as well as how the GB (second-order) and Lovelock (third-order) curvature terms actually help with using normal matter as opposed to exotic matter in the vicinity of the throat. Higher-order curvature corrections are used in wormholes with smaller throat radii, since the curvature near the throat is very large for such wormholes. The matter near the throat can be normal for the region $r_0 \leq r \leq r_{\max}$, where $r_{\max}$ depends on the Lovelock coefficients and the shape function. We obtain a larger radius (more region) with normal matter in the case of third-order Lovelock gravity with z negative coupling constant $\alpha_3$ as compared to the second-order Gauss–Bonnet wormholes. There is a lower limit on the throat radius $r_0$ imposed by the positivity of $\rho$ and $(\rho + p)$ in the case of Lovelock gravity but not in the case of Einstein gravity. The lower limit depends on the Lovelock coefficients, the dimensionality of the spacetime, and the shape function.

## 4.6 Wormholes in Einstein Dilaton Gauss–Bonnet (EdGB) Gravity

The authors in [71] reported stable wormhole solutions in EdGB gravity. Other authors reported in [72] that the same wormhole solution in EdGB gravity was indeed unstable. Since these two papers are excellent examples of stability analysis, in this section, we will review their key findings and methods.

In order to avoid using the required exotic matter necessary to support traversable wormholes, modified gravity theories were used. One such modified gravity theory is EdGB gravity. In this theory, the low-energy heterotic-string-theory-based effective action in four dimensions is given by [73, 74]

$$S = \frac{1}{16\pi} \int d^4x \sqrt{-g} \left( R - \frac{1}{2}\partial_\mu \phi \partial^\mu \phi + \alpha e^{-\gamma \phi} L^{(2)} \right), \tag{107}$$

where, in addition to the scalar curvature term $R$, we have a quadratic GB curvature term $L^{(2)}$ given by (48) and a scalar field term $\phi$ known as the dilaton field, with a coupling constant $\gamma$. $\alpha$ is a positive numerical constant given in terms of the Regge slope parameter.

In [71], the authors analyzed the stability of a spherically symmetric wormhole solution in EdGB gravity. With a coordinate transformation $r^2 = l^2 + r_0^2$, where $r_0$ is the radius at the throat, the spherically symmetric wormhole solution becomes pathology-free and is given by

$$ds^2 = -e^{2\nu(l)}dt^2 + f(l)dl^2 + (l^2 + r_0^2)(d\theta^2 + \sin^2\theta d\varphi^2). \tag{108}$$

In terms of the new coordinates, the expansion at the throat ($l = 0$) gives

$$f(l) = f_0 + f_1 l + ... \tag{109}$$

$$e^{2\nu(l)} = e^{2\nu_0}(1 + \nu_1 l) + ... \tag{110}$$

$$\phi(l) = \phi_0 + \phi_1 l + ..., \tag{111}$$

where $f_i$, $\nu_i$, and $\phi_i$ are constant coefficients, and all curvature invariants, including the GB term, remain finite for $l \to 0$. Next, we analyze the stability of the wormhole in terms of radial perturbations. The metric and dilaton functions depend on both $l$ and $t$. We then decompose the functions into unperturbed and perturbed parts. The perturbations are

$$\tilde{\nu}(l,t) = \nu(l) + \epsilon \delta\nu(l) e^{i\sigma t}, \tag{112}$$

$$\tilde{f}(l,t) = f(l) + \epsilon \delta f(l) e^{i\sigma t}, \tag{113}$$

$$\tilde{\phi}(l,t) = \phi(l) + \epsilon \delta\phi(l) e^{i\sigma t}. \tag{114}$$

$\sigma$ is defined in [71] as an unspecified eigenvalue that appears in their Ordinary Differential Equations (ODEs). Substituting these perturbations into the Einstein and Dilaton equations, as well as linearizing in $\epsilon$, we obtain a system of linear ODEs for the functions $\delta\nu(l)$, $\delta f(l)$, and $\delta\phi(l)$. Rearranging the ODEs, the authors arrived at a coupled second-order equation in $\delta\phi$ as follows:

$$(\delta\phi)'' + q_1(\delta\phi)' + (q_0 + q_\sigma \sigma^2)\delta\phi = 0, \tag{115}$$

where $q_0$, $q_1$, and $q_\sigma$ depend on the unperturbed solution. To be able to normalize, $\delta\phi \to 0$ as $l \to \infty$. At $l = 0$, $q_\sigma$ is bounded, and $q_0$ and $q_1$ diverge as $\frac{1}{l}$. To avoid this singularity, a transformation is made as follows:

$$\delta\phi = F(l)\psi(l), \tag{116}$$

where $F(l)$ satisfies the equation

$$\frac{F'}{F} = \frac{q_1(l)}{2}. \tag{117}$$

This implies that

$$\psi'' + Q_0 \psi + \sigma^2 q_\sigma \psi = 0, \tag{118}$$

where $Q_0 = -\frac{q_1'}{2} - \frac{q_1^2}{4} + q_0$ is bounded at $l = 0$. Then, $\psi \to 0$ as $l \to \infty$ and for $l = 0$. Solving these ODEs for several values of $\frac{\alpha}{r_0^2}$ and $f_0$, a solution exists only for certain values of the eigenvalues $\sigma^2$. The negative modes obtained for families of wormhole solutions indicate that the wormhole is unstable. In addition, there are positive modes in a region of parameter space where the EdGB wormhole solutions are linearly stable under radial perturbations.

In [72], the same wormhole was proved to be unstable for the following reason. In [71], the perturbation function is

$$\delta\phi(t, l) = A(l)\chi(t, l), \tag{119}$$

where the factor $A(l)$ diverges at the throat as $\frac{1}{l}$. Vanishing boundary conditions were imposed at the throat to obtain finite perturbation at the throat. This disconnects the regions of space on both sides of the throat for purely radial modes of perturbations. $\delta\phi$ is not a gauge invariant quantity. For general spherically symmetric perturbations of the wormhole, the gauge invariant quantity $\chi(t, l) \propto \gamma\delta\phi(t, l) - \frac{r^2}{l}\phi'(l)\delta r(t, l)$ satisfies the wavelike equation

$$\frac{\partial^2 \chi}{\partial t^2} - \frac{\partial^2 \chi}{\partial y^2} + V_{eff}(l)\chi(t, l) = 0, \tag{120}$$

where $V_{eff}$ is a finite effective potential. When $\chi$ is finite at the throat, $\delta\phi$ does not diverge unless the throat size is fixed by chosing $\delta r = 0$. Therefore, the wormhole is unstable with perturbation at the throat radius. The behavior of the dilaton field $\phi$ at the throat is a pure artifact of the chosen gauge and therefore can be safely ignored.

# 5 Other Wormhole Studies

## 5.1 Casimir Wormholes

Traversable wormhole solutions to Einstein's field equations require the existence of exotic matter (matter violating the weak, null, and dominant energy conditions). The most common type of exotic matter proposed is that which has negative energy density $\rho < 0$. Classically, negative energy density is thought to be impossible. However, in quantum field theory, negative energy densities can and do occur. The most famous example of this is the Casimir vacuum energy. The usual derivation of it follows quantizing the electromagnetic field between two neutral parallel plates and renormalizing the energy. This results in an energy density that is negative or less than the energy of the undisturbed vacuum:

$$\rho = \frac{-\hbar c \pi^2}{720 a^4}, \tag{121}$$

where $a$ is the separation between the plates. As a result, there has been much speculation and study undertaken into the possibility of using the Casimir vacuum energy to construct traversable wormholes.

This section will serve as an introduction to the basic concepts and results of attempting to use the Casimir vacuum energy to construct a traversable wormhole. Morris, Thorne, and Yurtsever [9, 75] were the first to publish the idea of utilizing the Casimir effect to support traversable wormholes. They noted that there were two distinct possibilities for utilizing the Casimir effect. The first is to consider the Casimir vacuum energy as is, i.e., with $a$ representing the fixed plate separation. The second is to promote the plate separation to the radial coordinate r.

The case of fixed plate separation was analyzed by Garattini [76], in which he studied the connection between the Casimir energy and absurdly benign traversable wormholes. He utilized the semiclassical Einstein field equations:

$$G_{\mu\nu} = \langle T_{\mu\nu} \rangle, \tag{122}$$

where $\langle T_{\mu\nu} \rangle$ is the renormalized quantum expectation value of the stress–energy tensor, and $G_{\mu\nu}$ is the classical Einstein tensor. The justification for this equation was first given by Hawking [77] in his derivation for particle creation by black holes, in which he argued that quantum gravity contributions to the Einstein field equations can be assumed negligible above the Planck length. The same assumption is made here. Upon solving the Einstein field equations using the metric for a static spherically symmetric traversable wormhole, defined by Morris and Thorne, the shape function is found to be

$$b(r) = r_0 - \frac{\pi^3}{720 a^4} \left( \frac{\hbar G}{c^3} \right) (r^3 - r_0^3), \tag{123}$$

where $r_0$ is the throat radius. It is clear from this expression that this spacetime is not asymptotically flat, and in fact, it is asymptotically de Sitter. As stated in [76], the Casimir vacuum energy can then be viewed as a cosmological constant. However, in close proximity to the throat, the shape function can be approximated as

$$b(r) \approx r_0 \left( 1 - \frac{r_0 l_p^2 \pi^3}{90 a^4} (r - r_0) \right), \tag{124}$$

where it has been assumed that $r$ is very close to $r_0$, and $l_p$ is the Planck length. This shape function has the same form as the shape function for "absurdly benign traversable wormholes" defined by Morris and Thorne [9], namely

$$b(r) = r_0 \left( 1 - \frac{(r - r_0)}{d} \right)^2 ; \Phi(r) = 0; r_0 \leq r \leq r_0 + d, \tag{125}$$

$$b(r) = 0; \Phi(r) = 0; r > r_0 + d, \tag{126}$$

where $d$ is given by (128), and exotic matter is confined in the region ($r_0 \leq r \leq r_0 + d$).

In order to make the connection between the Casimir shape function and this shape function, the zero tidal force (ZTF) condition must be imposed, where $\Phi(r) = 0$. The leading order for this shape function close to the throat is

$$b(r) = r_0 \left( 1 - 2 \frac{(r - r_0)}{d} \right), \tag{127}$$

which gives us the required connection between these two shape functions:

$$d = \frac{90 a^4}{r_0 l_p^2 \pi^3}. \tag{128}$$

The absurdly benign traversable wormhole is defined so that the exotic matter is confined close to the throat ($d$ is small). In order to fit this requirement, there are two possibilities: the first is $r_0 > 10^{34}$ m, and the second is $a \sim 10^{-15}$ m for a wormhole with $r_0 \sim 10^{10}$ m. Neither of these conditions are physical. It is noted that the identification made for the relationship between $d$ and $a$ is a physically meaningless assumption. Therefore, Garattini attempted to refine and reduce the throat radius by examining various examples of inhomogeneous equations of state, and he managed to establish a throat radius of $r_0 \sim 10^{17}$ m for a plate separation of 1 nm. However, these results still present a nonphysical result, and they point to the likelihood that the semiclassical Einstein field equations may be ill suited for the description of Casimir traversable wormholes.

Promoting the plate separation to radial coordinates was discussed in [78] as follows. Once again, semiclassical quantum gravity was used but now with the Casimir vacuum energy density given by

$$\rho = \frac{-\hbar c \pi^2}{720 r^4} \tag{129}$$

where $r$ is the radial coordinate. This gives the following shape function

$$b(r) = r_0 - \frac{r_1^2}{r_0} + \frac{r_1^2}{r} \tag{130}$$

Here, $r_1^2 = \frac{\pi^3}{l_p^2}$. First, observe that this cannot be transformed into an absurdly benign traversable wormhole. In fact, attempting to impose the ZTF condition produces disastrous results. Instead, we use the Einstein field equation and the shape function to solve the redshift function. We also use the EOS $p_r = \omega \rho$. In solving the redshift function, it is necessary to impose the condition $\omega r_1^2 - r_0^2 = 0$. This condition is required for the existence of the wormhole, since if $\omega r_1^2 - r_0^2 > 0$, we obtain a black hole. With this condition, we obtain the following shape function and redshift function, respectively,

$$b(r) = \left(1 - \frac{1}{\omega}\right) r_0 + \frac{r_0^2}{\omega r} \tag{131}$$

and

$$\Phi(r) = \frac{1}{2}(\omega - 1) \ln\left(\frac{r\omega}{r\omega + r_0}\right). \tag{132}$$

These result in a wormhole of Planckian size.

We now return to the question of using the semiclassical Einstein field equations. As stated above, Hawking originally justified the use of this approximation in the derivation of Hawking radiation. In this scenario, it makes sense that quantum gravitational effects can be ignored in the Einstein tensor, since Hawking never considered Planckian-sized black holes, as the black holes would be expected to have evaporated before reaching this size. So, any such effects could be assumed to be hidden by the formation of the event horizon.

When considering the use of the Casimir effect in constructing traversable wormholes, we find that we might not be able to ignore quantum gravity. First, as we have just seen above, promoting the plate separation to a radial coordinate results in a Planckian-sized wormhole negating Hawking's assumption of not having to deal with any system of this size. For the fixed plate separation, we encountered solutions that produce wormholes with throat radii on the order of the Sun's radius for plate separation on the order of nanometers and smaller. This makes no physical sense, as the Casimir vacuum energy is confined within the plates and would have no way of interacting with a body of such size. Lastly, we do not have any event horizon by definition and therefore cannot assume that quantum gravity effects are hidden away. So far, no one has made an attempt to incorporate any such quantum gravity effects into the analysis of Casimir wormholes, and it is therefore left as the subject for future publications. However, other studies have been done on the subject of Casimir traversable wormholes. For example, Garattini [79] studied the effects of an electric charge on a Casimir wormhole. He found that the throat radius becomes directly dependent on the strength of the electric charge, meaning that there exists the possibility of creating a throat larger than the Planck length. However, it should be noted that, in his analysis, it was found that there exists a radius at which the energy density becomes positive. It is not discussed whether this negates the existence of the wormhole.

Casimir wormholes have also been studied in general D dimensions [80]. It has been shown that, even in these scenarios, the radius is found to be on the order of the Planck length or a nonphysical size. It is worth reiterating that these studies have been constructed primarily using the semiclassical Einstein field equations and therefore do not include any possible effects of quantum gravity, which is a fact that is most likely the source of many of the nonphysical results.

## 5.2  Thin-Shell Wormholes in Modified Gravity

There is a set of wormhole solutions in modified gravity theories that use special techniques to satisfy the energy conditions and thus remove the need for exotic matter. We summarize the analysis of these wormholes here. In [81], the authors derived asymptotically flat traversable wormhole solutions that satisfy the NEC in a quadratic form of the gravitational hybrid metric-Palatani gravity theory. Their solution has an interior part with a nonexotic perfect fluid near the throat, an exterior Schwarzschild solution, and a double gravitational layer thin shell at the junction hypersurface $\Sigma$ between the interior and exterior solution.

The generalized hybrid metric-Palatani gravity theory has the action

$$S = \frac{1}{2\kappa^2} \int_\Omega \sqrt{-g} f(R, \mathcal{R}) d^4 x + \int_\Omega \sqrt{-g} \mathscr{L}_m d^4 x. \tag{133}$$

Here, $\kappa^2 \equiv \frac{8\pi G}{c^4}$, $\Omega$ is the spacetime manifold in which a set of coordinates $x^a$ are defined, $g$ is the determinant of the metric $g_{ab}$, and $f(R, \mathcal{R})$ is a well-behaved function of $R$ and $\mathcal{R}$. $R \equiv g^{ab} R_{ab}$ is the Ricci scalar, $\mathcal{R} \equiv g^{ab} \mathcal{R}_{ab}$ is the Palatini scalar curvature, and $\mathcal{R}_{ab}$ is the Palatini Ricci tensor written in terms of an independent connection $\Gamma^c_{ab}$. $\mathscr{L}_m$ is the matter Lagrangian density minimally coupled to the metric $g_{ab}$. The junction conditions at the separation hypersurface $\Sigma$ were defined to be

$$[h_{\alpha\beta}] = 0, \tag{134}$$

$$[K] = 0, \tag{135}$$

$$[R] = 0, \tag{136}$$

$$[\mathcal{R}] = 0, \tag{137}$$

$$f_{R\mathcal{R}}n^a[\partial_a R] + f_{\mathcal{R}\mathcal{R}}n^a[\partial_a \mathcal{R}] = 0, \tag{138}$$

and

$$\epsilon\delta^\beta_\alpha n^c[\partial_c R]\left(f_{RR} - \frac{f_{R\mathcal{R}}^2}{f_{\mathcal{R}\mathcal{R}}}\right) - (f_R + f_\mathcal{R})\epsilon\left[K^\beta_\alpha\right] = 8\pi S^\beta_\alpha. \tag{139}$$

where $h_{\alpha\beta} = e^a_\alpha e^b_\beta g_{ab}$ is the induced metric at the hypersurface $\Sigma$, $K_{\alpha\beta} = e^a_\alpha e^b_\beta \nabla_a n_b$ is the extrinsic curvature, $K = K^\alpha_\alpha$ is the trace of the extrinsic curvature, and $S^\beta_\alpha$ is the stress–energy tensor of the thin shell arising at the hypersurface $\Sigma$. It was shown in [81] that the general set of junction conditions shown above can be simplified for particular forms of $f(R,\mathcal{R})$. In particular, they selected an $f(R,\mathcal{R})$ that is quadratic in $R$ and linear in $\mathcal{R}$. For this $f(R,\mathcal{R})$, the junction conditions $[R] = 0$ and $[\mathcal{R}] = 0$ are not mandatory conditions anymore. This gives rise to additional terms in the stress–energy tensor $S_{ab}$ of the thin shell, which then forms a double layer thin-shell distribution at the junction hypersurface $\Sigma$.

Unlike the general case [82], in which the solutions obtained using scalar–tensor representation are scarce, the simplified set of junction conditions in [81] implies that it is possible to obtain numerous solutions for a wide variety of metrics and actions. There is also the advantage that asymptotic flatness can be preserved in this case, unlike the general case where the NEC can be guaranteed only for the asymptotically flat AdS spacetime.

In [83], traversable wormholes with double layer thin shells in quadratic gravity were analyzed, where $f(R) \equiv R + \alpha R^2$, in which R is the Ricci scalar. The NEC is satisfied in this case at the throat, as well as the whole wormhole interior. However, the NEC is not satisfied for the double layer stress–energy distribution component at the thin shell.

In [84], traversable wormhole solutions were analyzed in $f(R,\mathcal{T})$ gravity theory for a linear model, i.e., $f(R,\mathcal{T}) = R + \gamma\mathcal{T}$, where $\mathcal{T} = T_{ab}T^{ab}$, and $T_{ab}$ is the energy–momentum tensor. It was shown that there are a large set of wormhole solutions in which the matter field satisfies all the energy conditions (NEC, WEC, SEC, and DEC). Since the field equations are quadratic in the matter quantities $\rho$, $p_r$, and $p_t$, as well as complex, they used an analytical recursive algorithm to extract the nonexotic wormhole solutions. The solutions obtained were not naturally localized, so the junction conditions were derived for this theory. It was proven that a matching between two spacetimes must always be smooth and does not allow thin shells at the boundary. Traversable localized static and spherically symmetric wormhole solutions satisfying all energy conditions were obtained by matching the interior wormhole spacetime to an exterior vacuum Schwarzschild solution.

## 5.3 Dark Matter Halo Wormholes

It is well known that an exotic type of matter with negative energy density can stabilize wormholes. Dark matter halos are vast, invisible regions of space that surround galaxies. They are composed of dark matter [85], a substance that does not emit, absorb, or reflect light, so they cannot be directly detected. Their presence is inferred through their gravitational effect on visible matter such as stars and gas clouds. In recent studies of galaxy formation, it was found that every galaxy forms within a dark matter halo [86]. The formation and growth of galaxies over time is related to the growth of the halos in which they form.

In recent years, a few research groups have been investigating traversable wormholes supported by the dark matter halo [87–90]—both in GR and modified theories of gravity such as f(Q,T). Dark matter halo wormholes are usually studied using the Milky Way galaxy (MWG) halo profiles, pseudothermal, Navarro–Frenk–White (NFW) models I and II, and Universal Rotation Curves (URCs). A sample investigation of dark matter profiles in the Milky Way galaxy can be found in [91]. In [90], they used the "Einasto dark matter density profile" to produce suitable redshift and shape functions. The NEC is violated for redshift functions $\Phi = C$ and $\Phi = \frac{\alpha}{r}$, and the shape function satisfies the flare-out condition. The anisotropic dark matter content within the wormhole creates the appropriate environment to stabilize the wormhole structure by violating the NEC. The global monopole charge $\eta$ [92–94] plays an important role in the violation of the NEC. The probability of violation of the NEC decreases for an increasing value of $\eta$, and so it is important to minimize the value of $\eta$.

In [95], wormholes supported by galactic halos have been investigated in *4D* EGB gravity. This analysis was done for three different dark matter profiles, namely URC, NFW, and Scalar Field Dark Matter (SFDM). The NEC was violated at the neighborhood of the wormhole throat. The Gauss–Bonnet coefficient $\alpha$ has an influence on the NEC. $\alpha > 0$ gives negative energy. The contribution to the violation increases with $\alpha$. The SEC is also violated for each of the DM profiles.

### 5.4 Wormholes in Nonlocal Theories of Gravity (NLGs)

There is a class of nonlocal integral kernel theories of gravity where the inverse of the d'Alembert operator in the gravitational action is taken into account. These are called nonlocal theories of gravity (NLGs). A spherically symmetric MT wormhole solution satisfying NLG field equations was studied in [96]. Linear and exponential NLG correction terms were selected due to the existence of Noether symmetry [97]. It was found that the nonlocal gravity contributions allow for the stability and traversability of a wormhole without considering exotic matter.

## 6 Summary and Discussion

In this review paper, our goal is to review stable, traversable wormholes in $f(R)$-like gravity theories. We started with a historic development of ideas about wormholes in the Introduction section and then provided some motivation for this paper and the need for an equation of state (EOS) in Section 1.1. We then concluded the Introduction with a discussion of the astronomical observational signatures of natural wormholes in Section 1.2. In Section 2.1, we started with a brief discussion of each modified gravity theory such as $f(R)$, $f(R, \mathscr{T})$, Lovelock, EGB, Brans–Dicke, and KK theory, and we followed with a brief discussion of nonmetricity theories such as $f(Q)$ and $f(Q, T)$. We ended Section 2.1 with a review of $f(R, L_m)$ theory and its applications. In Sections 2.2 and 2.3, we gave a more detailed treatment of $f(R)$ gravity theory and Lovelock gravity theory by deriving the EOM from the action for the respective theory. In Section 3, we discussed the MT wormhole in greater detail, since it is the standard traversable wormhole geometry used in most of the studies of wormhole in modified gravity theories, as seen in the literature. In Section 4, we reviewed the study of wormholes in f(R) gravity and Lovelock gravity based on the flare-out condition and energy condition violation requirements. We concluded Section 4 with a detailed review of stability analysis under a perturbation using EdGB gravity as an example. In Section 5, we took the reader on a tour of other wormholes such as Casimir wormholes, thin-shell wormholes in modified gravity, dark matter halo wormholes, and wormholes in nonlocal theories of gravity.

In the examples that we discuss, it is possible to have stable wormholes without the use of exotic matter in many of the modified gravity theories. In some cases, it is shown that the amount of exotic matter that is needed can be minimized. In their analyses, the authors varied the redshift function, shape function, the type of fluid used, the equation of state, and the primary function used in the corresponding modified gravity theory.

The key takeaways for wormholes in $f(R)$ gravity theory are as follows: In GR, violation of the NEC is required for static traversable wormholes. In the $f(R)$ theory of gravity, modified field equations are obtained by varying the modified action with respect to the metric. Using these modified field equations, we require that matter (stress–energy tensor) threading the wormhole satisfy the NEC, and the required violation of the NEC can be enabled by the total stress–energy tensor, which includes higher-order curvature terms. The higher-order curvature terms, interpreted as a gravitational fluid, support the nonstandard wormhole geometries. A constant redshift function is assumed in many of these analyses to reduce the complexity of the calculations. It is possible to use a variable redshift function as well. Based on the review of papers related to wormholes in Lovelock gravity, we have the following key takeaways: Higher-order curvature terms become useful for the analysis of wormholes with smaller throat radius. There exists a lower limit for the throat radius in Lovelock gravity imposed by the requirement that $\rho$ and $(\rho + p)$ be positive. There is no such limit in Einstein's gravity. The radius of the region with normal matter is higher for wormholes in third-order Lovelock gravity, with a negative coupling constant ($\alpha_3$), compared to wormholes in second-order Gauss–Bonnet gravity.

In Section 5, we discussed "Other Wormholes" such as Casimir, thin-shell, dark matter halo, and nonlocal gravity wormholes. In Casimir wormholes, the required negative energy density for a stable traversable wormhole is provided by the Casimir effect. Casimir wormholes are currently being studied in several modified gravity theories. Thin-shell wormholes use special techniques to satisfy the energy conditions and thus remove the need for exotic matter. For example, there are asymptotically flat traversable wormhole solutions that satisfy the NEC in the quadratic form of the gravitational hybrid metric-Palatani gravity. These solution have an interior portion with a nonexotic perfect fluid near the throat, an exterior Schwarzschild solution, and a double gravitational layer thin shell at the junction hypersurface $\Sigma$ between the interior and exterior solutions. In dark matter halo wormholes, the dark matter content within the wormhole creates the appropriate environment to stablize the wormhole structure by violating the NEC. In NLG wormholes, it was found that the nonlocal gravity contributions allow for the stability and traversability of a wormhole, without considering exotic matter.